\documentstyle[12pt]{article}
\begin{document}
\def\bea{\begin{eqnarray}}
\def\eea{\end{eqnarray}}
\title{\bf {
 The Cardy-Verlinde formula and entropy of Topological Kerr-Newman black holes in de
Sitter spaces }}
\author{$^{1,2,3}$M.R. Setare\thanks{%
E-mail: rezakord@yahoo.com} and $^{4}$M.B.Altaie\thanks{%
E-mail: maltaie@yu.edu.jo} \\
{$^{1}$Physics Dept. Inst. for Studies in Theo. Physics and
Mathematics(IPM) }\\
{P. O. Box 19395-5531, Tehran, IRAN }\\
and \\
{$^{2}$Department of Science, Physics group, Kordestan University,
Sanandeg,
Iran}\\
and \\
{\ $^{3}$Department of Physics, Sharif University of Technology,
Tehran,
Iran }\\
{$^{4}$Department of Physics, Yarmouk University, Irbid-Jordan}}

\maketitle
\begin{abstract}
In this paper we show that the entropy of cosmological horizon in
4-dimensional Topological Kerr-Newman-de Sitter  spaces can be
described by the Cardy-Verlinde formula, which is supposed to be
an entropy formula of conformal field theory in any dimension.
Furthermore, we find that the entropy of black hole horizon can
also be rewritten in terms of the Cardy-Verlinde formula for these
black holes in de Sitter spaces, if we use the definition due to
Abbott and Deser for conserved charges in asymptotically de Sitter
spaces. Such result presume a well-defined dS/CFT correspondence,
which has not yet attained the credibility of its AdS analogue.

 \end{abstract}
\newpage

 \section{Introduction}
Holography is believed to be one of the fundamental principles of
the true quantum theory of gravity\cite{{HOL},{RAP}}. An
explicitly calculable example of holography is the much-studied
AdS/CFT correspondence. Unfortunately, it seems that we live in a
universe with a positive cosmological constant which will look
like de Sitter space-time in the far future. Therefore, we should
try to understand quantum gravity or string theory in de Sitter
space preferably in a holographic way. Of course, physics in de
Sitter space is interesting even without its connection to the
real world; de Sitter entropy and temperature have always been
mysterious aspects of quantum gravity\cite{GH}.\\
While string theory successfully has addressed the problem of
entropy for black holes, dS entropy remains a mystery. One reason
is that the finite entropy seems to suggest that the Hilbert space
of quantum gravity for asymptotically de Sitter space is finite
dimensional, \cite{{Banks:2000fe},{Witten:2001kn}}.
 Another, related, reason is that the horizon and entropy in
de Sitter space have an obvious observer dependence. For a black
hole in flat space (or even in AdS) we can take the point of view
of an outside observer who can assign a unique entropy to the
black hole. The problem of \ what an observer venturing inside the
black hole experiences, is much more tricky and has not been given
a satisfactory answer within string theory. While the idea of
black hole complementarity provides useful clues, \cite
{Susskind}, rigorous calculations are still limited to the
perspective of the outside observer. In de Sitter space there is
no way to escape the problem of the observer dependent entropy.
This contributes to the difficulty of de Sitter space.\\
More recently, it has been proposed that defined in a manner
analogous to the AdS/CFT correspondence,  quantum gravity in a de
Sitter (dS) space is dual to a certain
 Euclidean  CFT living on a spacelike boundary of the
dS space~\cite{Strom} (see also earlier works
\cite{Hull}-\cite{Bala}). Following the proposal, some
investigations on the dS space have been carried out
recently~\cite{Mazu}-\cite{Ogus}. According to the dS/CFT
correspondence, it might be expected that as the case of AdS black
holes~\cite{Witten2}, the thermodynamics of cosmological horizon
in asymptotically dS spaces can be identified with that of a
certain Euclidean CFT residing on a spacelike boundary of the
asymptotically dS spaces.\\
One of the remarkable outcomes of the AdS/CFT and dS/CFT
correspondence has been the generalization of Cardy's formula
(Cardy-Verlinde formula) for arbitrary dimensionality, as well as
a variety AdS and dS backgrounds. In this paper, we will show that
the entropy of cosmological horizon in the 4-dimensional
Topological Kerr-Newman-de Sitter  spaces (TKNdS) can also be
rewritten in the form of Cardy-Verlinde formula. We then show that
if one uses the Abbott and Deser (AD) prescription \cite{AD}, the
entropy of black hole horizons in dS spaces can also be expressed
by the Cardy-Verlinde formula \cite{cai1}. In a previous paper
\cite{set}, we have shown that the entropy of cosmological horizon
in Topological Reissner-Nordstr\"om- de Sitter spaces in arbitrary
dimension can be described by the Cardy-Verlinde formula. Each of
these cases is found to have interesting implications in the
context of the proposed correspondence. The Cardy-Verlinde formula
in 4-dimensional Kerr-Newman-de Sitter has been studied previously
in \cite{jing} (the dS/CFT correspondence have considered for the
three-dimensional Kerr-de Sitter space already at \cite{mu2}), In
the KNdS case the cosmological horizon geometry is spherical, also
the black hole have two event horizon, but in the TKNdS case the
cosmological horizon geometry is spherical, flat and hyperbolic
for k=1,0,-1, respectively, also for the case k=0,-1 the black
hole have not event horizon. In the other hand the entropy of such
spaces come from both cosmological and event horizon, then absence
of event horizon or existanc of extra cosmological horizon change
the result for the total entropy.

\section{Topological Kerr-Newman-de Sitter Black Holes}
The line element of TKNdS black holes in 4-dimension is given by
\begin{eqnarray}
ds^{2} &=&-\frac{\Delta _{r}}{\rho ^{2}}\left(dt-\frac{a}{\Xi
}\sin ^{2}\theta d\phi \right)^{2}+\frac{\rho ^{2}}{\Delta
_{r}}dr^{2}+\frac{\rho ^{2}}{\Delta
_{\theta }}d\theta ^{2} \nonumber  \\
&&+\frac{\Delta _{\theta }\sin ^{2}\theta }{\rho ^{2}}\left[a
dt-\frac{ (r^{2}+a^{2})}{\Xi }d\phi \right]^{2},  \label{kdsmet}
\end{eqnarray}
where
\begin{eqnarray}
\Delta _{r} &=&(r^{2}+a^{2})\left(k-\frac{r^{2}}{l^{2}}\right)
-2Mr+q^2,  \nonumber \\
\Delta _{\theta } &=&1+\frac{a^{2}\cos ^{2}\theta}{ l^{2}},
  \nonumber \\
\Xi &=&1+\frac{a^{2}}{l^{2}},  \nonumber \\
\rho ^{2} &=&r^{2}+a^{2}\cos ^{2}\theta .  \label{kdelt}
\end{eqnarray}
here the parameters $M$, $a$, and $q$ are associated with the
mass, angular momentum, and electric charge parameters of the
space-time, respectively. The topological metric Eq.(1) will only
solve the Einstein equations if k=1, which is the spherical
topology. In fact when $k=1$, the metric Eq.(1) is just the
Kerr-Newman -de Sitter solution. Three real roots of the equation
$\Delta _{r}=0$, are the location of three horizon, the largest on
is the cosmological horizon $r_{c}$, the smallest is the inner
horizon of black hole, the one in between is the outer
horizon $r_{b}$ of the black hole.\\
If we want the $k=0,-1$ cases solve the Einstein equations, then
we must set $sin\theta \rightarrow \theta$, and $sin\theta
\rightarrow sinh\theta$ respectively \cite{man1}-\cite{man4}. When
$k=0$ or $k=-1$, there is only one positive real root of $\Delta
_{r}$, and this locates the position of cosmological horizon
$r_{c}$. When $q=0$, $a=0$ and $M\rightarrow -M$ the metric
Eq.(1)is the TdS (Topological de Sitter) solution
\cite{{cai93},{med}}, which have a cosmological horizon and a
naked singularity, for this type of solution, the Cardy-Verlinde
formula also work well.\\
Here we review the BBM prescription \cite{BBM} for computing the
conserved quantities of asymptotically de Sitter spacetimes
briefly. In a theory of gravity, mass is a measure of how much a
metric deviates near infinity from its natural vacuum behavior;
i.e, mass measures the warping of space. Inspired by the analogous
reasoning in AdS space \cite{{by},{b}} one can construct a
divergence-free Euclidean quasilocal stress tensor in de Sitter
space by the response of the action to variation of the boundary
metric, in 4-dimensional spacetime we have
\begin{eqnarray}
T^{\mu \nu} &=& {2 \over \sqrt{h}} { \delta I \over \delta h_{\mu
\nu}} = \ \  {1 \over 8\pi G} \left[ K^{\mu\nu} - K \, h^{\mu\nu}
+ {2 \over l} \, h^{\mu\nu} +l  \,G^{\mu\nu} \right] ,
 \label{stressminus}
\end{eqnarray}
where $h^{\mu\nu}$ is the metric induced on surfaces of fixed
time, $K_{\mu\nu}$, $K$ are respectively extrinsic curvature and
its trace, $G^{\mu\nu}$ is the Einstein tensor of the boundary
geometry. To compute the mass and other conserved quantities, one
can write the metric $h^{\mu\nu}$ in the following form
\begin{equation}
    h_{\mu\nu} \, dx^{\mu} \, dx^{\nu } =
       N_{\rho}^{2} \, d\rho^{2} +
       \sigma_{ab}\, (d\phi^a + N_\Sigma^a \, d\rho) \,
               (d\phi^b + N_\Sigma^b \, d\rho)
       \label{boundmet}
\end{equation}
where the $\phi^{a}$ are angular variables parametrizing closed
surfaces around the origin. When there is a Killing vector field
$\xi^{\mu}$ on the boundary, then the conserved charge associated
to $\xi^{\mu}$ can be written as \cite{{by},{b}}
\begin{equation}
   Q =  \oint_{\Sigma}  d^{2}\phi \,\sqrt{\sigma } \,
   n^{\mu}\xi^{\mu} \,T_{\mu\nu}
   \label{chargedef}
\end{equation}
where $n^{\mu}$ is the unit normal vector on the boundary,
$\sigma$ is the determinant of the metric $\sigma_{ab}$. Therefore
the mass of an asymptotically de Sitter space in 4-dimension is as
\begin{equation}
    M =
    \oint_{\Sigma}  d^{2}\phi \,\sqrt{ \sigma } \, N_{\rho} \,
\epsilon
    ~~~~~;~~~~~ \epsilon \equiv
    n^{\mu}n^{\nu} \,
    T_{\mu\nu} \, ,
    \label{massdef}
\end{equation}
where Killing vector normalized as $\xi^{\mu} = N_{\rho} n^{\mu}$.
Using this prescription \cite{BBM}, the gravitational mass,
subtracted the anomalous Casimir energy, of the 4-dimensional
TKNdS solution is
\begin{equation}
E=\frac{-M}{\Xi}. \label{bbmass}
\end{equation}
Where the parameter $M$ can be obtain from equation
$\Delta_{r}=0$. On this basis, the following relation for the
gravitational mass can be obtained
\begin{equation}
E=\frac{-M}{\Xi} =\frac{(r_c^2+a^2)(r_c^2-k l^2)-q^2 l^2}{2\Xi r_c
l^2}. \label{kmass}
\end{equation}
The Hawking temperature of the cosmological horizon given by
\begin{equation}
T_{c}=\frac{-1}{4\pi}\frac{\Delta
'_{r}(r_{c})}{(r_c^2+a^2)}=\frac{3r_c^4+r_c^2(a^2-k
l^2)+(ka^2+q^2)l^2}{4\pi r_c l^2(r_c^2+a^2)}. \label{tem}
\end{equation}
The entropy associated with the cosmological horizon can be
calculated as
\begin{equation}
S_{c}=\frac{\pi(r_{c}^{2}+a^{2})}{\Xi}.
 \label{ent}
\end{equation}
The angular velocity of the cosmological horizon is given by
\begin{equation}
\Omega_{c}=\frac{-a\Xi}{(r_{c}^{2}+a^{2})}.
 \label{ang}
\end{equation}
The angular momentum $J_{c}$, the electric charge $Q$, and the
electric potentials $\phi_{qc}$ and $\phi_{qc0}$ are given by
\begin{eqnarray}
  & & {\mathcal{J}}_c=\frac{M a}{\Xi^2}, \nonumber \\
 & & Q =\frac{q}{\Xi}, \nonumber  \\
 & & \Phi_{qc}=-\frac{q r_c}{r_c^2+a^2}, \nonumber  \\
 & & \Phi_{qc0}=-\frac{q}{r_c},
 \label{ctherm}
\end{eqnarray}
The obtained above quantities of the cosmological horizon satisfy
the first law of thermodynamics
\begin{equation}
dE=T_cdS_c+\Omega_c d{\mathcal{J}}_c+(\Phi_{qc}+\Phi_{qc0}) dQ .
\label{Flth}
\end{equation}

 Using the
Eqs.(\ref{ent},\ref{ctherm}) for cosmological horizon entropy,
angular momentum and charge, also equation $\Delta_{r}(r_{c})=0$,
we can obtain the mtric parameters $M$, $a$, $q$ as a function of
$S_{c}$, ${\mathcal{J}}_c$  and $Q$, after that we can write $E$
as a function of these thermodynamical quantities
$E(S_{c},{\mathcal{J}}_c,Q)$ (see \cite{{cal},{dehkh}}). Then one
can define the quantities conjugate to $S_{c}$, ${\mathcal{J}}_c$
and $Q$, as
\begin{equation}
T_c=\left( \frac{\partial E}{\partial S_c}\right) _{J_c,_Q},\ \
\Omega_c =\left( \frac{\partial E}{\partial {J}_c}\right)
_{S_c,Q},\ \ \Phi_{qc} =
\left( \frac{\partial E%
}{\partial Q}\right) _{S_c,J_c}\\\Phi_{qc0} =lim_{a\rightarrow 0}
\left( \frac{\partial E%
}{\partial Q}\right) _{S_c,J_c}\\  \label{Dsmar}
\end{equation}
 Making use of the fact
that the metric for the boundary CFT can be determined only up to
a conformal factor, we rescale the boundary metric for the CFT to
be the following form
\begin{equation}
ds_{CFT}^2=\lim_{r \rightarrow \infty}\frac{R^2}{r^2}ds^2 ,
\label{euniverse}
\end{equation}
Then the thermodynamic relations between the boundary CFT and the
bulk TKNdS are given by
\begin{equation}
E_{CFT}=\frac{l}{R}E,\hspace{0.07 cm}T_{CFT}=\frac{l}{R}T
,\hspace{0.07 cm}J_{CFT}=\frac{l}{R}J,\hspace{0.07
cm}\phi_{CFT}=\frac{l}{R}\phi,\hspace{0.07
cm}\phi_{0CFT}=\frac{l}{R}\phi_{0}, \label{CFT}
\end{equation}
 The Casimir energy $E_c$,
defined as $E_C =(n+1)
E-n(T_cS_c+J_c\Omega_c+Q/2\phi_{qc}+Q/2\phi_{qc0})$ , and $n=2$ in
this case, is found to be
\begin{equation}
E_C=-\frac{k(r_c^2+a^2)l }{R \Xi r_c}, \label{ckecas01}
\end{equation}
in KNdS space case \cite{jing} the Casimir energy $E_c$ is always
negative, but in TKNdS space case Casimir energy can be positive,
negative or vanishing depending on the choice of $k$.
 Thus we can see that the entropy
 Eq.(\ref{ent})of the cosmological horizon can be rewritten as
 \begin{equation}
 S=\frac{2\pi R}{n}\sqrt{|\frac{E_{c}}{k}|(2(E-E_q)-E_c)},\label{careq}
\end{equation}
where
\begin{equation}
E_q = \frac{1}{2}\phi_{c0} Q.\label{qeq}
\end{equation}
We note that  the entropy expression (\ref{careq}) has a similar
form as the
case of TRNdS black holes \cite{set}.\\
 For the black hole
horizon, which there is only for the case $k=1$ associated
thermodynamic quantities are
\begin{equation}
T_{b}=\frac{1}{4\pi}\frac{\Delta
'_{r}(r_{b})}{(r_b^2+a^2)}=-\frac{3r_b^4+r_b^2(a^2-l^2)+(a^2+q^2)l^2}{4\pi
r_b l^2(r_b^2+a^2)}. \label{tem2}
\end{equation}
\begin{equation}
S_{b}=\frac{\pi(r_{b}^{2}+a^{2})}{\Xi}.
 \label{ent2}
\end{equation}
\begin{equation}
\Omega_{b}=\frac{a\Xi}{(r_{b}^{2}+a^{2})}.
 \label{ang2}
\end{equation}
\begin{equation}
   {\mathcal{J}}_b=\frac{M a}{\Xi^2},
\end{equation}
\begin{equation}
Q =\frac{q}{\Xi},
\end{equation}
\begin{equation}
\Phi_{qb}=\frac{q r_b}{r_b^2+a^2},
 \end{equation}
\begin{equation}
\Phi_{qb0}=\frac{q}{r_b}.
\end{equation}
Now if we uses the BBM mass Ee.(\ref{bbmass}) the black hole
horizon entropy cannot be expressed by a form like Cardy-Verlinde
formula \cite{cai93}. The other way for computing conserved
quantities of asymptotically de Sitter space is Abbott and Deser
(AD) prescription \cite{AD}. According to this prescription, the
gravitational mass of asymptotically de Sitter space coincides
with the ADM mass in asymptotically flat space, when the
cosmological constant goes to zero. Using the AD prescription for
calculating conserved quantities the black hole horizon entropy of
TKNdS space can be expressed in term of Cardy-Verlind formula
\cite{cai1}. The AD mass of TKNdS solution can be expressed in
terms of black hole horizon radius $r_b$, $a$ and charge $q$,
\begin{equation}
  E'=\frac{M}{\Xi} =\frac{(r_b^2+a^2)(r_b^2-l^2)-q^2 l^2}{2\Xi r_b l^2}.
\label{kmass2}
\end{equation}
The obtained above quantities of the black hole horizon also
satisfy the first law of thermodynamics as
\begin{equation}
dE'=T_bdS_b+\Omega_b d{\mathcal{J}}_b+(\Phi_{qb}+\Phi_{qb0}) dQ .
\label{Flth1}
\end{equation}

The thermodynamics quantities of the CFT must be rescaled by a
factor $\frac{l}{R}$ similar to the pervious case. In this case,
the Casimir energy, defined as
 $ E'_C
=(n+1) E' -n(T_bS_b+J_b\Omega_b+Q/2\phi_{qb}+Q/2\phi_{qb0})$, is
\begin{equation}
 E'_C =\frac{(r_b^2+a^2)l }{R \Xi r_b}, \label{ckecas02}
\end{equation}
and the black hole entropy $ S_{b}$ can be rewritten as
\begin{equation}
S_{b}=\frac{2\pi
R}{n}\sqrt{E'_{C}|(2(E'-E'_q)-E'_C)|},\label{careq2}
\end{equation}
where
\begin{equation}
E'_q =\frac{1}{2}\phi_{qb0} Q.\label{qeq2}
\end{equation}
which is the energy of electromagnetic field outside the black
hole horizon. Thus we demonstrate that the black hole horizon
entropy of TKNdS solution can be expressed in a form as the
Cardy-Verlinde formula. However, if one uses the BBM mass
Eq.(\ref{kmass}) the black hole horizon entropy $S_{b}$ cannot be
expressed by a form like the Cardy-Verlinde formula. Our result is
in favour of the dS/CFT correspondence.
\section{Conclusion}
The Cardy-Verlinde formula recently proposed by  Verlinde
\cite{Verl}, relates the entropy of a  certain CFT to its total
energy and Casimir energy in arbitrary dimensions. In the spirit
of dS/CFT correspondence, this formula has been shown to hold
exactly for the cases of dS Schwarzschild, ds topological, ds
Reissner-Nordstr\"om , dS Kerr, and dS Kerr-Newman  black holes.
In this paper we have further checked the Cardy-Verlinde formula
with topological Kerr-Newman  de Sitter black hole. \\
It is well-known that there is no black hole solution whose event
horizon is not sphere, in de Sitter background although there are
such solutions in anti-de Sitter background, then in TKNdS space
for the case k=0,-1 the black hole have not event horizon, however
the cosmological horizon geometry is spherical, flat and
hyperbolic for k=1,0,-1, respectively. As we have shown there
exist two different temperature and entropy associated with the
cosmological horizon and black hole horizon, in TKNdS spacetimes.
If the temperatures of the black hole and cosmological horizon are
equal, then the entropy of system is the sum of entropies of
cosmological and black hole horizons. The geometric features of
black hole temperature and entropy seem to imply that the black
hole thermodynamics is closely related to nontrivial topological
structure of spacetime. In \cite{cai2} Cai, et al in order to
relate the entropy with Euler characteristic $\chi$ of the
corresponding Euclidean manifolds have been presented the
following relation
\begin{equation}
S=\frac{\chi_{1}A_{BH}}{8}+\frac{\chi_{2}A_{CH}}{8},
\end{equation}
in which the Euler number of the manifolds divided into two parts,
one first part comes from the black hole horizon and the second
part come from the cosmological horizon (see also
\cite{{teit},{gib},{lib}}). If one uses the BBM mass of the
asymptotically dS spaces, the black hole horizon entropy cannot be
expressed by a form like the Cardy-Verlinde formula\cite{cai93}.
In this paper, we have found that if one uses the AD prescription
to calculate conserved charges of asymptotically dS spaces, the
TKNdS black hole horizon entropy can also be rewritten in a form
of Cardy-Verlinde formula, which indicates that the thermodynamics
of black hole horizon in dS spaces can be also described by a
certain CFT. Our result is also reminiscent of the Carlip's
claim~\cite{Carlip}(to see new formulation which is free of
inconsistencies encountered in Carlip's ref.\cite{mu})  that for
black holes in any dimension the Bekenstein-Hawking entropy can be
reproduced using the Cardy formula~\cite{Cardy}. Also we have
shown that the Casimir energy for cosmological horizon in TKNdS
space case can be positive, negative or vanishing depending on the
choice of $k$, by contrast the Casimir energy for cosmological
horizon in KNdS space is always negative \cite{jing}.

  \vspace{3mm}

\section*{Acknowledgement }
We would like to thank Prof. Robert Mann for useful comments and
suggestions.

\end{document}